\documentclass[runningheads]{llncs}

\usepackage{url}
\usepackage{makeidx}
\usepackage{graphicx} 
\usepackage{newtxmath}
\usepackage[bottom]{footmisc}
\usepackage{amsmath}
\usepackage{hyperref}
\usepackage{dcolumn}
\usepackage{color}
\usepackage{float}
\usepackage{bm}

\begin{document}
\title{The electrodynamic origin of the wave-particle duality}
%
\titlerunning{Electromagnetic pilot waves}  %
%
\author{\'Alvaro Garc\'ia L\'opez\inst{1}\orcidID{0000-0002-2485-4639}}
\authorrunning{A. G. L\'{o}pez.}
%
\institute{Universidad Rey Juan Carlos, M\'ostoles s/n 28933, Madrid, Spain,\\
\email{alvaro.lopez@urjc.es}}

\maketitle              
\begin{abstract}
A derivation of pilot waves from electrodynamic self-interactions is presented. For this purpose, we abandon the current paradigm that describes electrodynamic bodies as point masses. Beginning with the Li\'enard-Wiechert potentials, and assuming that inertia has an electromagnetic origin, the equation of motion of a nonlinear time-delayed oscillator is obtained. We analyze the response of the uniform motion of the electromagnetic charged extended particle to small perturbations, showing that very violent oscillations are unleashed as a result. The frequency of these oscillations is intimately related to the \emph{zitterbewegung} frequency appearing in Dirac's relativistic wave equation. Finally, we compute the self-energy of the particle. Apart from the rest and the kinetic energy, we uncover a new contribution presenting the same fundamental physical constants that appear in the quantum potential.

\keywords{nonlinear dynamics \and chaos \and delay differential equations \and electrodynamics \and retarded potentials \and pilot waves \and quantum mechanics}.
\\

\url{https://doi.org/10.1007/978-3-030-99792-2_88}

\end{abstract}
\section{Introduction}

Recently developed models of silicon droplets have shown deep connections between quantum mechanical systems and classic hydrodynamics, allowing nonlinear dynamicists to grasp how the complex motion of a quantum particle can be \cite{pro06,for10}. More specifically, these macroscopic systems describe the unpredictable dynamics of walking droplets as the result of a feedback interaction between the bouncing particle and the waves that it produces when it strikes the surface of a fluctuating medium underneath, beyond Faraday's threshold. Fortunately, and contrary to quantum mechanical models, these hydrodynamic analogs are investigated in terms of understandable and firmly established principles of chaos theory and nonlinear dynamical systems. 

Although the pilot wave dynamics of silicon droplets has been proposed as a candidate to comparatively investigate quantum systems, a specific homologous mechanism that can give rise to the wave-particle duality in the microscopic realm has not been rigorously developed until very recently \cite{onan20}. In the present paper we provide strong evidence suggesting that the wave-particle duality has its basis in the theory of classical electromagnetism. For this purpose, we show that extended charged bodies can self-interact when they are accelerated. A certain region of the particle can emit radiation, which later on affects a different region of the same particle. This phenomenon introduces a time-delay in the self-force of the extended body.

Consequently, the description of the dynamics of charged bodies must be posed by means of retarded differential equations. As it is well-known, the solutions to these differential equations frequently present limit cycle behaviour as a consequence of the Andronov-Hopf bifurcation \cite{gha09,sta21}. The feedback interaction of radiative and Coulombian fields among different charged parts of the particle can trigger a fast oscillation, destabilizig its uniform motion. These fields produce dissipation and antidamping as a consequence of radiation reaction. In the thermodynamic context of open systems, such a periodic motion has been recently referred as a self-oscillation \cite{jen13}. In this manner, we show that the wave-particle duality is just an immediate consequence of the self-oscillation of extended electrodynamic bodies, which can be regarded as dissipative structures.

\section{Electrodynamics of an extended body}
We model the electrodynamics of an extended charged body by using the Lagrangian density of Maxwell's theory of electrodynamics with sources. This density is written as
\begin{equation}
\mathcal{L} =- \frac{1}{4 \mu_{0}} F_{\mu \nu}F^{\mu \nu}-A_{\mu}J^{\mu},
\label{eq:1}
\end{equation}
where $J^{\mu}$ denotes the four-current density representing the sources, and $F^{\mu \nu}$ is the Faraday tensor. Then, Maxwell's equations can be derived in covariant form from the previous action by differentiation, yielding
\begin{equation}
\partial_{\mu} F^{\mu \nu}=\mu_{0} J^\nu.
\label{eq:2}
\end{equation}

To describe the dynamics of the source of charge, we express the four-density as $J^\mu=\rho_0 U^{\mu}$, where the density of charge $\rho_0$ in the proper frame and the four-velocity $U^{\mu}$ have been introduced. The charge density $\rho$ in some inertial reference frame can be related to the density in the proper frame by using the Lorentz factor $\gamma$, through the relation $\rho=\gamma \rho_0$. Then, the four-current is simply written as $J^{\mu}=(\rho c,J)$, with $J=\rho v_s$ the euclidean current density. Here we are considering a rigid distribution of charge. This assumption allows us to write the charge density at time $t$ as $\rho(x,t)=\rho(x-x_{s}(t))$, where the vector $x_{s}(t)$ represents the position of the particle's centre of mass at time $t$. Under these assumptions, Maxwell's equations can be solved in terms of retarded potentials, which allow to generally derive the fields by means of Jefimenko's equations. Using the fact that $F_{\mu\nu}=\partial_{\mu}A_{\nu}-\partial_{\nu}A_{\mu}$, the four-potential in the Lorenz gauge can then be written as
\begin{equation}
A^{\mu}(x,t)=\frac{1}{4 \pi \epsilon_{0}c^2} \int  \dfrac{J^{\mu}(x,t_r)}{|x-x'|} d^3x',
\label{eq:3}
\end{equation}
where we have introduced the retarded time $t_r=t-|x-x'|/c$. The analysis presented ahead considers a very simple rigid charged distribution comprised of two point particles at a fixed distance. Therefore, in order to compute the self-force, we can simply use the \emph{Li\'enard-Wiechert potential}. This potential is the solution to a charged point particle, whose charge density can be represented by means of the Dirac delta distribution, in the form $\rho(x,t) = q\delta(x-x_s(t))$. In the Lorenz gauge we have the solution 
\begin{equation}
A^{\mu}(x,t)=\frac{q}{4 \pi \epsilon_{0}c}\left(\frac{v^{\mu}}{(1- n_s \cdot \beta_s)|x-x_s|} \right)_{t=t_r},
\label{eq:4}
\end{equation}
where the relative speed of the source $\beta_s(t)=v_s(t)/c$ has been defined, together with the time-like four-vector $v_s^{\mu}=(c,v_s(t))$. Finally, the unit vector $n_s(x,t)=(x-x_s(t))/|x-x_s(t)|$ has also been introduced. It points to the spatial point $x$, where we want to compute the value of the fields. Its application point is located at the position $x_s(t)$, where the charge was placed at the retarded time $t=t_r$.

\subsection{A model of an electron}

\begin{figure}
\centering
\includegraphics[width=0.8\linewidth,height=0.47\linewidth]{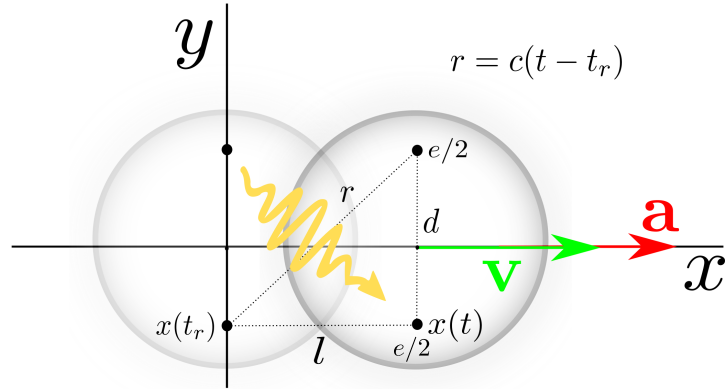}
\caption{An electron is shown at the retarded time $t_{r}$ and at the present time $t$. It consists of two point charges placed along the y-axis at a constant distance $d$. From $t_{r}$ to $t$, the particle accelerates and advances a distance $l$ along the x-axis. A field perturbation is shown emerging from the upper point at the retarded time (yellow photon). Later on, this perturbation exerts a force on the second point at the lower side of the body. In this manner, an extended corpuscle can feel itself in the past. The speed $v$ and the acceleration $a$ of the particle are represented in green and red, for clearness.}
\label{fig:1}
\end{figure}

In the preset work we consider the charge density $\rho(x,t)=-(e/2)\delta(x-x_s(t))\delta(z)(\delta(y+d/2)+\delta(y-d/2))$. It is the most simple model of an electron, represented as an ``extended" electrodynamic body, as shown in Fig.~\ref{fig:1}. This charge density represents a body formed by two points with charge $-e/2$ placed at a fixed distance $d$ in the y-axis, which moves along the x-axis. We restrict the motion to transversal displacements to simplify our study, since the non-conservative character of electrodynamics with sources makes the computations very entangled, due to the fact that the Li\'enard-Wiechert potential is retarded in time. By restricting ourselves to a one-dimensional translational motion, we avoid the more complicated three-dimensional problem, including self-torques. This fudamental model has been designed in previous works to underpin the use of the \emph{Abraham-Lorentz} force and also to study the contribution of electromagnetic mass to inertia \cite{gri89}. We use this elementary model hereafter, which suffices to explain the physical mechanism leading to the wave-particle duality. 

\subsection{The self-interaction}

As shown in Fig.~\ref{fig:1}, an extended electrodynamic body can interact with itself. This kind of interaction is frequently called a \emph{self-interaction} \cite{gri89}. The upper charged point exerts an electromagnetic force on the lower point a short time later. The self-force appears because electromagnetic waves can travel between the two points of the electrodynamic body. Apart from a force of inertia, a term of damping and a restoring elastic force, it can induce the self-exicted motion of the particle due to a radiation reaction force. We now compute the electric field produced by the upper point at the lower point of the body. If we utilize the Li\'enard-Wiechert potential of a point particle, we obtain
\begin{equation}
\bm{E}=\frac{q}{8\pi \epsilon_{0}}\frac{r}{(\bm{r}\cdot\bm{u})^3}\left(\bm{u}(1-\beta^{2})+\frac{1}{c^2}\bm{r}\times (\bm{u}\times\bm{a})\right),
\label{eq:5}
\end{equation}
where the relative position between the two points at different times is $\bm{r}(t_{r})$, the normalized velocity is $\bm{\beta}(t_{r})=\bm{v}(t_{r})/c$, the acceleration $\bm{a}(t_{r})$ and the vector $\bm{u}=\bm{\hat{r}}-\bm{\beta}$ has been introduced. We highlight that these kinematic variables depend on the retarded time $t_{r}=t-r/c$. This time-delay appears because electromagnetic field perturbations travel with limited velocity in spacetime, according to the principle of \emph{causality}. This limitation puts a constraint $r=c(t-t_{r})$ on the self-interaction, assigning a specific event in the past light cone from which the signals coming from one point of the particle can affect the remaining point. 

According to Fig.~\ref{fig:1}, we can write the position as $\bm{r}=l\bm{\hat{x}}+d\bm{\hat{y}}$, the velocity as $\bm{\beta}=v/c\bm{\hat{x}}$ and the acceleration vector as $\bm{a}=a \bm{\hat{x}}$. These relations allow to compute the vector $\bm{u}$ as
$\bm{u}=(l-r\beta)/r\bm{\hat{x}}+d/r\bm{\hat{y}}$. Then, making use of the identity $r^{2}=(x(t)-x(t_{r}))^2+d^2$, the following inner product $\bm{r}\cdot\bm{u}=r-l\beta$ results. Regarding the radiative component of the field, we need to expand the double cross-product in the form $\bm{r}\times(r\bm{u}\times\bm{a})=-d^2 a \bm{\hat{x}}+d a l \bm{\hat{y}}$. The total self-force on the center of mass of the particle can be written as $\bm{F}_{\rm{self}}=-e\bm{E}$, where the symmetry of the arrangement has been taken into consideration. Because of the rigidity of the charge density, we recall that the the magnetic attractive forces and the electric repulsive forces all cancel each other along the y-axis. The resulting force that the particle exerts on itself is
\begin{equation}
\bm{F}_{\rm{self}}=\frac{e^2}{8\pi \epsilon_{0}} \frac{1}{(r-l \beta)^3} \left( (l-r \beta)(1-\beta^{2})-\frac{d^2}{c^2}a \right) \bm{\hat{x}}.
\label{eq:7}
\end{equation}

\subsection{Time-delayed equation of motion}

Following the tradition, we could now invoke Newton's second law of mechanics. For a non-relativistic particle it is written as $\bm{F}_{\rm{self}} = m \bm{a}$, with $m$ the electron's bare mechanical mass. However, it has been shown in recent works that the self-force can be expanded by using a Taylor series of the time-delay $r/c$ \cite{onan20}. Among the infinite linear and nonlinear terms that contribute to the self-force, the most well-known are the Lorentz-Abraham force, which consists of a linear term proportional to the jerk ($\dot{a}$) of the particle, and the term of inertia, which is proportional to the acceleration. This term dominates over all other terms in the limit of very small $d/c$, what allows to approximate the self-force as $\bm{F}_{\rm{self}} = -m_e \bm{a}$ for non-relativistic velocities. To this end, we simply define the electromagnetic mass as $m_{e} c^2 = e^2/16 \pi \epsilon_{0} d$, where Einstein's mass-energy relation has been obviously used. 

In the present work we are assuming that the rest mass of the electron comes entirely from its electrostatic energy, so that its bare mass can be made equal to zero. For if mass is not a fundamental property of particles, but just energy, all the mass in our model must come from the electrostatic energy of the two point charges. Then, if we use Sommerfeld's relation for the fine structure constant, the rest mass of the electron is
\begin{equation}
m_{e}=\frac{\hbar \alpha}{4 d c}.
\label{eq:8}
\end{equation}
Using this relation, we can approximate a electron radius of $r_{e} = d / 2 = 0.35~\rm{fm}$. Naturally, this value is closely related to the the electron's classical radius. 

Therefore, this approach does not artificially introduce bare  mechanical inertia in the theory of classical electromagnetism. Instead, we use the \emph{principle of D'Alembert} that, in the approximation of macroscopic objects, leads to Newton's second law. Thus classical mechanics should be considered an emergent theory resulting from averaging magnitudes over large numbers of electrodynamic bodies. As a corollary, we predict that the gravitational force between two particles have an electrodynamic origin as well. Consequently, we propose to replace Newton's second law by the static problem
\begin{equation}
\bm{F}_{\rm{ext}}+ \bm{F}_{\rm{self}} = 0.
\label{eq:9}
\end{equation} 
As long as we can approximate $\bm{F}_{\rm{self}} = -m_e \bm{a}$ in the macroscopic limit, we see that Newton's second law naturally arises from Maxwell's dynamical theory of the electromagnetic field. The force of inertia reveals in this way as an electromagnetic force of \emph{self-induction}, coming from the interior of the body as a consequence of Faraday's law. This statement opposes to Mach's principle, which tries to justify the origin of inertial forces on external distant masses. 

To study the ``free'' particle, we can settle the external forces to zero, thus we have the simple law of motion $\bm{F}_{\rm{self}} = 0$. Its solution describes the geodesic motion of the electrodynamic body in the same way as in the theory of general relativity, for example. The differential equation of motion is
\begin{equation}
\frac{d^2}{c^2}a(t_{r})+\frac{r}{c} \left(1-\frac{v^2(t_{r})}{c^2} \right) v(t_{r})+\left(1-\frac{v^2(t_{r})}{c^2} \right) \left(x(t_r)-x(t)\right)=0.
\label{eq:10}
\end{equation}
The difficulty with this state-dependent \emph{delayed differential equation} \cite{sie17} is that most kinematic variables are specified at the retarded time $t_r = t - r / c$. If we translate them to the present time $t \rightarrow t + r /c$, we obtain
\begin{equation}
a(t)+\frac{r}{d} \frac{c}{d} \left(1- \frac{v^2(t)}{c^2} \right) v(t)+\left(\frac{c}{d}\right)^2 \left(1-\frac{v^2(t)}{c^2} \right) \left(x(t)-x \left( t+\frac{r}{c} \right) \right)=0.
\label{eq:11}
\end{equation}

This differential equation clearly evokes a nonlinear oscillator \cite{jen13}. We can identify a term of Newtonian inertia and a typical linear oscillating term representing an elastic restoring force. But we can see two more nonlinear contributions, as well. Firstly, the contribution appearing in the second term acts as a nonlinear damping force, producing the system's dissipation. Secondly, the advanced potential produces a non-conservative force of antidamping. 

The frequency of oscillation can be approximated as $\omega_{0} = c/d$, what yields a value of the period $T_{0}=4 \pi r_{e}/c=1.18 \times 10^{-22} \rm{s}$, if we use the  electron's classical radius. Thus the electron oscillates very fast describing a deterministic motion. However, this motion resembles a stochastic motion at large enough time scales. This jittery dynamics and the specific value of its period are very familiar to quantum mechanical theorists. They are closely related to the trembling motion appearing in Dirac's wave equation equation for relativistic particles, commonly known as \emph{zitterbewegung}. 

Importantly, we notice that the time-delay and the damping term involve an \emph{arrow of time}. Irreversibility is inherent to non-conservative dynamical systems presenting limit cycle behavior. It is also conventional in time-delayed systems, whose trajectories are not specified by some initial conditions, but rely on the complete knowledge of functions describing part of their previous history. Of course, this non-conservative dynamics only appears when we try to describe the motion of the particle solely, without reference to the dynamical fields. 

In summary, fundamental particles can be considered as open dissipative structures. They are locally active and operate far from equilibrium by taking and releasing electromagnetic energy to their surroundings. The dissipative nature of classical electrodynamics with sources becomes manifest by the fact that a Lagrangian density for the motion of the particle cannot be written by using the traditional minimal coupling, as it is frequently done in quantum particle physics.

\section{Stability analysis}

Now we prove that transversal motion at constant speed is not stable. Consequently, self-oscillatory motion is the only possibility, irrespective of the periodicity of this nonlinear oscillation. For this purpose, we consider the differential equation (\ref{eq:11}) and pose it in the phase space canonical variables. We obtain
\begin{align}
&\dot{x} =v, \nonumber \\
&\dot{v} =-\frac{c}{d}\frac{r}{d}\left(1-\frac{v^2}{c^2} \right)v-\left(\frac{c}{d}\right)^2\left(1-\frac{v^2}{c^2} \right)\left(x-x_{\tau} \right). \label{eq:12} 
\end{align}
The variable $x_{\tau}$ has been introduced. It represents the electron's position evaluated at time $t+\tau$, where we recall that $\tau=r/c$. Consider that motion at constant speed $\beta c$ is feasible. Since $x(t) = v t$, we also have $x(t + r / c) = vt + vr /c$, what yields $x - x_{\tau} = -vr /c$. If we now substitute in Eq.~(\ref{eq:12}) we get
\begin{align}
&\dot{x} =v,\nonumber \\
&\dot{v} =-\frac{c}{d}\frac{r}{d}\left(1-\frac{v^2}{c^2} \right)v+\frac{c}{d}\frac{r}{d}\left(1-\frac{v^2}{c^2} \right)v=0. \label{eq:13} 
\end{align}

Therefore, every uniform motion is an invariant solution of our delayed dynamical system. We now prove that these solutions are unstable as well. For this purpose, we compute the variational equations
\begin{align}
&\delta\dot{x} =\delta v,\nonumber \\
&\delta \dot{v} = -\frac{c}{d}\frac{\delta r}{d}\left(1-\frac{v^2}{c^2}\right)v-\frac{c}{d}\frac{r}{d}\left(1-\frac{v^2}{c^2}\right)\delta v+\frac{c}{d}\frac{r}{d}\frac{2 v^2}{c^2}\delta v-\nonumber \\ 
&~~~~~~-\frac{c}{d}\frac{r}{d}\frac{2 v^2}{c^2} \delta v-\left(\frac{c}{d}\right)^2\left(1-\frac{v^2}{c^2} \right)\left(\delta x-\delta x_{\tau} \right). 
\label{eq:14} 
\end{align}
We follow by computing $\delta r$ when $\dot{v} = 0$, with $v = \beta c$. For this purpose, we use the relation $r^{2}=(x(t)-x(t_{r}))^2+d^2$ and Eq.~(\ref{eq:10}). If we combine these two equations we can derive polynomial of degree two in $r$. If we also introduce the Lorentz factor $\gamma = (1 - \beta^{2})^{-1/2}$, the solution to this polynomial yields
\begin{equation}
r=\gamma d \sqrt{1+\gamma^6\dot{\beta}^2 \left(\frac{d}{c}\right)^2}+\gamma^{4} c \beta \dot{\beta}\left(\frac{d}{c}\right)^2,
\label{eq:15}
\end{equation}
Recall, the speed and the acceleration appearing in Eq.~\ref{eq:15} are evaluated at time $t_{r}$. Interestingly, the \emph{time-delay} becomes a function of the kinematic variables. When the particle increases its speed, the self-force is originated at an earlier past time, because the light cone of the particle evolves with its motion. Evaluating the Eq.~(\ref{eq:15}) at time $t$, the computation of the variations of $r$ can be done immediately, yielding
\begin{equation}
\delta r(t)=\gamma^4 \beta \left(\frac{d}{c}\right)^2 \delta \dot{v}(t)+d \delta \gamma(t).
\label{eq:16}
\end{equation}
Grouping terms and using the equation $r = \gamma d$ for $\dot{v} = 0$, we obtain the variational equation
\begin{align}
&\delta\dot{x} =\delta v,\nonumber \\
&\delta \dot{v} \gamma^2 = -\frac{c}{d}\gamma \delta v-\left(\frac{c}{d}\right)^2 \left(1-\beta^2 \right) \left(\delta x-\delta x_{\tau} \right).\label{eq:17} 
\end{align}

\begin{figure}
\centering
\includegraphics[width=0.65\linewidth,height=0.57\linewidth]{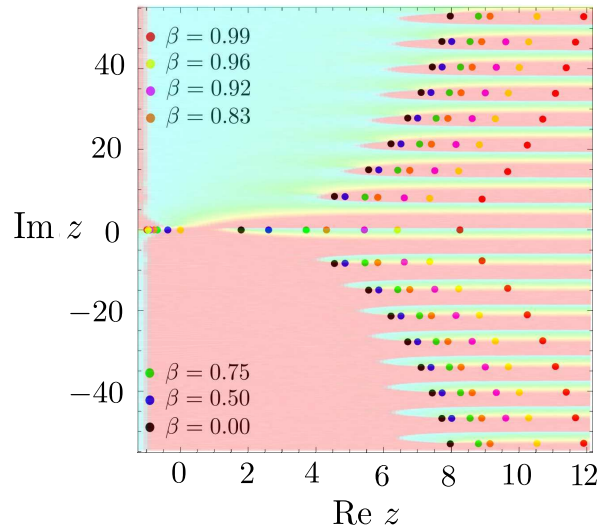}
\caption{The complex function $f(z)=z^2+z+(1-\beta^2)(1-e^z)$ and its roots are computed for seven different values of the velocity, with the aid of Newton-Raphson method. Since $z=\gamma d/c$, we get the spectrum of eigenfrequencies of the self-oscillation, which can be approximated as $\omega_{n} \propto n c/\gamma d$. The background corresponds to a representation of the function for $\beta=0$, using a domain coloring technique.}
\label{fig:3}
\end{figure}

If we consider exponential solutions $ \delta x = A e^{\lambda t}$, the characteristic equation of the dynamical system (\ref{eq:17}) can be found. It reads
\begin{equation}
\mu^2 +\mu+(1-\beta^2)(1-e^{\mu})=0,
\label{eq:18}
\end{equation}
where the variable $\mu =\lambda \gamma d/c$ has been defined. With the exception of one eigenvalue, the solutions to this equation always present a positive real part, what guarantees their unstability for all values of $\beta$. As depicted in Fig.~\ref{fig:3}, this analytical statement is confirmed by numerical simulation. An infinite spectrum of \emph{eigenvalues} of the frequency is obtained, with the following quantization rule
\begin{equation}
\omega_{n}=\eta_{n} \frac{c}{\gamma d}.
\label{eq:19}
\end{equation}
The appearence of the $\gamma$ factor is due to the Lorentz boost, which produces a time dilation. The form factor $\eta_{n}$ is characteristic of the geometry of the body. In summary, we have proved mathematically the existence of oscillatory motion in our dynamical system for any value of the relative velocity $\beta$. Importantly, we would like to mention that this instability depends on the shape of the particle. If the geometry of the electrodynamic body is switched from oblate to prolate, a Hopf bifurcation occurs \cite{sta21}.

\section{The quantum potential}

We conclude the present paper by deriving the relativistic kinetic energy and the quantum potential from the Li\'{e}nard-Wiechert potential. The insertion of Eq.~\eqref{eq:15} into the equation $r^2=l^2+d^2$ allows to exactly obtain $l$ as a function of the relative speed $\beta$ and the relative acceleration $\dot{\beta}$. The result yields the equation
\begin{equation}
l=\sqrt{\gamma^2c^2\beta^2\left(\frac{d}{c}\right)^2+\gamma^{8}c^{2}\dot{\beta}^{2}(1+\beta^{2})\left(\frac{d}{c}\right)^4+2 c^2 \gamma^{5} \beta \dot{\beta}\left(\frac{d}{c}\right)^3 \sqrt{1+\gamma^6\dot{\beta}^2\left(\frac{d}{c}\right)^2}}.
\label{eq:20}
\end{equation}

Now we denote the self-energy of the particle as $E$, which we define as the energy of non-dissipative origin needed to build the charge and achieve its particular dynamical state of motion. As it is well-known, the magnetic fields perform no work, and because dissipated energy is being disregarded, we focus on the curl-free part of the electric field. Bearing in mind these considerations, the potential energy $E$ of the electrodynamic particle can be defined by means of the Li\'enard-Wiechert potential as $E=-e c A^{0}/2$, what yields
\begin{equation}
E=\frac{e^2}{16 \pi \epsilon_{0}}\frac{1}{\bm{r} \cdot \bm{u}}.
\label{eq:21}
\end{equation}
Using previous relations, this equation is written as
\begin{equation}
E=\frac{\hbar \alpha c}{4(r-l \beta)}.
\label{eq:22}
\end{equation}
If we replace the Eqs.~(\ref{eq:15}) and (\ref{eq:20}) and expand the resulting self-potential in powers of $d/c$, we obtain the following Taylor series expansion
\begin{equation}
E=\gamma \frac{\hbar \alpha c}{4d}-\gamma^7 \frac{a^2}{2 c^2}\frac{\hbar \alpha}{4}\left(\frac{d}{c}\right)+\gamma^{13} \frac{3 a^4}{8 c^4}\frac{\hbar \alpha}{4}\left(\frac{d}{c}\right)^3-\gamma^{19} \frac{5 a^6}{16 c^6}\frac{\hbar \alpha}{4}\left(\frac{d}{c}\right)^5+...
\label{eq:23}
\end{equation}
Again, we assume the idea that mass and inertia have a total electromagnetic origin. Therefore, the size of the particle can be written using Eq.~(\ref{eq:8}) as $d=\hbar \alpha/4 m_{e} c$. Noticeably, mass $m_{e}$ is proportional to $\hbar$. This relation implies that any kind of energy or canonical momentum can be written as proportional to Planck's constant. Furthermore, if the speed of the particle is related to the group velocity of the pilot wave, then it seems reasonable to consider that the relation $p=\hbar k$ holds. This introduces De Broglie's relation connecting the velocity of the particle and the wavelength of the electromagnetic pilot wave. Substitution of Eq.~\eqref{eq:8} in Eq.~\eqref{eq:23} yields
\begin{equation}
E=\gamma m_{e}c^2+Q,
\label{eq:24}
\end{equation}
where we have introduced the potential
\begin{equation}
Q=\frac{\hbar^2}{2 m_{e}} \frac{\alpha^2}{8 d^2} \gamma \sum_{n=1}^{\infty}{\frac{(-1)^n(2n-1)!!}{2^n n!}\gamma^{6n} \frac{a^{2n}}{c^{2n}}}\left(\frac{d}{c}\right)^{2n}.
\label{eq:25}
\end{equation}
We detect two well differentiated terms in Eq.~\eqref{eq:24}. The former corresponds to the famous relativistic equation representing the energy of a particle in the theory of special relativity. It contains the rest energy of the particle together with its kinetic energy. Importantly, we stress that these two magnitudes are not fundamental and correspond to plain electrodynamic energy. Additionaly, apart from the rest mass and the kinetic energy, the new potential energy $Q$ has appeared. By using a quadrature related to the coefficients in Eq.~\eqref{eq:25}, the series can be computed and one last integration yields \cite{onan20} the expression
\begin{equation}
Q=-\frac{\hbar^2}{2 m_{e}} \frac{\alpha^2}{8 d^2} \gamma \left(1-1 / \sqrt{1+\gamma^6 \dot{\beta}^2 \left(d/c\right)^2} \right).
\label{eq:26}
\end{equation}
This new contribution vanishes for uniform motion. The Lorentz factor prevents the particle from traveling at velocities equal or above the speed of light. The constant term $\hbar^2/2m_{e}$ preceding this potential is identical to the quantum potential appearing in Bohmian mechanics \cite{boh52}, which can be written as $Q=-(\hbar^2/2 m_{e})\nabla^2 R/R$. This potential can not be derived from a Hamiltonian including an external source of potential, and involves a self-organising process produced by the internal electromagnetic field \cite{onan20}. The quantum potential entails an interpretation of classical electrodynamic phenomena in terms of an emergent hydrodynamic theory \cite{sch54,lop212}, which overcomes the representation of complicated internal self-interactions by using the concept of quantum pressure.

In theory, once the dynamical system approaches its asymptotic limit set, a functional relation between the position of the particle in the configuration space and its acceleration can be provided. This relation can be replaced in $Q(x,t)$, allowing to compute the function $R(x,t)$. Then, we can pose the Hamilton-Jacobi equation. If the particle is also subjected to the influence of a newtonian external potential $V(x,t)$ and its average velocity is not relativistic, such an equation reads
\begin{equation}
\frac{\partial S}{\partial t}+\frac{1}{2 m_{e}}(\nabla S)^2+Q+V=0.
\label{eq:27}
\end{equation}

After solving the previous equations, and making use of the information about the particle's trajectory, the wave function can be built using the polar expression $\psi(x,t)=R(x,t)\exp\left(i S(x,t)/\hbar \right)$. Importantly, we deduce from these relations that the wave function is not an ordinary probabilistic entity, but a \emph{real} physical field \cite{boh52} related to external and internal electrodynamic fields. These fields describe the pilot wave of the particle, which can produce well-known physical phenomena, as for example interference and diffraction.

A conservative approximation of the quantum potential has been derived in recent works, connecting it to typical potentials that break fundamental symmetries \cite{onan20}. In particular, it has been claimed that the quantum potential can produce the symmetry breaking of the Lorentz group. Importantly, we recall that \emph{symmetry breaking} is an essential feature in the study of nonlinear dynamics \cite{lop22}. 

\section{Conclusions and Discussion}

The dynamics of an extended electrodynamic body has major similarities with the motion of silicon droplets found in many experiments during the past decade. In the present model, quantum waves have their origin in the self-oscillation of the electron, produced by the feedback interaction of the particle with its own electromagnetic field. This self-interaction enforces a pilot wave travelling with the corpuscle. Therefore, the wave-particle duality is immediately explained, since quantum waves appear naturally as perturbations of the dynamic electromagnetic fields. These self-interactions and their concomitant forces of recoil, perhaps together with external \emph{zero-point fluctuations} \cite{del14}, can prevent the collapse of the hydrogen atom \cite{raj04}. 

The fact that electromagnetic mass allows to derive the exact relativistic kinetic energy analytically from the electrodynamic potentials strongly suggests that inertial mass is not a fundamental concept in physics, but an emergent one. This conclusion points towards the fact that gravitational mass is also a redundant idea in fundamental physics, and that the force of gravity has an electromagnetic origin as well. In this perspective, an electrodynamic theory of the gravitational force would also explain in simple terms the principle of equivalence. The equality of gravitational mass and inertial mass would then be explained because of their common origin in the electromagnetic force. Importantly, our finding that Newton's second law can be deduced from classical electrodynamics shows that classical mechanics is an emergent theory based on classical electromagnetism. Just in the same way as thermodynamics laws result from averaging mechanical properties over large number of ensembles. Consequently, equations in which the concept of mass appears as an elementary parameter, as it occurs with the Sch\"odinger or the Dirac equations, should not be considered fundamental in physics.

Finally, the model presented in this work is not very realistic because it considers a rigid charge density, which is structurally unstable. More reasonably, it is expected that fundamental particles can arise from self-confined fields as a consequence of the rotation of the fields and their electromagnetic stress, stabilizing the electron. This idea suggests that particles are \emph{electromagnetic solitons} \cite{fab12,ran95}, perhaps arising in the context of the Einstein-Maxwell equations. Then, \emph{zitterbewegung} could be regarded from the point of view of a field theory as a quasi-breather solution.

Theories including more complex lagrangian densities with other general relativistic invariants or even nonlinear electrodynamic fields \cite{gul21} might also allow to understand fundamental particles as compact field configurations. Then, the electric charge could be explained as the \emph{topological charge} of the solitary wave, and not as a fundamental parameter. Anyway, no theory of elementary particles can be considered a fundamental theory as long as it does not conform to the principle of general covariance. Only this principle allows adopting any reference frame to describe the motion of dynamical fields, dispensing with the metaphysical concept of absolute spacetime.

%
%
%
\bibliographystyle{splncs04}
%

\end{document}